 \documentstyle[epsf]{article}

\newcommand{\iras}{IRAS\,04210+0400}
\newcommand{\kms}{{\rm\,km\,s}$^{-1}$}
\newcommand{\oiii}{[O{\sc iii}]-5007\AA~}

\newcommand{\etal}{{et al.}}
\newcommand{\rc}{r_1}

\newcommand{\kpc}{{\rm \,kpc\,}}
\newcommand{\hubble}{{$\rm H_o=75\,km\,s^{-1}\,Mpc^{-1}$}}

\begin{document}

\begin{centering}
{\Large \bf Jets and the emission-line spiral structure in \iras} \\
\vspace*{3mm}
     
{\large W. Steffen$^{1}$, A.J. Holloway$^{1}$, A. Pedlar$^{2}$} \\
\vspace*{3mm}

{\it
$^1$Department of Physics and Astronomy, University of Manchester,
Schuster Laboratory, Oxford Road, Manchester M13 9PL \\
$^2$Nuffield Radio Astronomy Laboratories, University of Manchester,
Jodrell Bank, Macclesfield, Cheshire SK11 9DL, UK}

\end{centering}

\begin{abstract}
We examine models in which jets are responsible for the formation of
the emission-line spiral structure in \iras.  The kiloparsec-scale
radio lobes in this active galaxy appear to be related to its extended
emission-line spiral structure.  The radio structure consists mainly
of extended symmetrically bent, FR\,I-type lobes, which follow the
emission-line spiral structure at their inner edge. In the central
region of the galaxy a double radio source is observed with a
separation of approximately 1\,arcsec between its components, which
are extremely well aligned with the hotspot from which the southern
lobe expands outwards. Hill \etal\ (1988) suggested a model for the
emission-line spiral structure invoking compressed interstellar
matter, which is dragged away from the original jet path by the
rotating ambient medium. From consideration of the propagation speed
of the jets and the transverse ram pressure exerted by the rotating
environment, we exclude this scenario as a possible origin of the
spiral structure. We favour a model in which the jets themselves are
bent by the rotating interstellar medium and possibly follows the
emission-line spiral arms. We present fits of the model to the
observed optical spiral structure. High sensitivity radio observations
will be required to decide on the nature of the peculiar spiral
structure in IRAS\,04210+0400.

\end{abstract}


\section{Introduction}

The nature of the spiral structure of disk galaxies has been a widely
discussed issue in extragalactic astronomy since their discovery.
Several mechanisms have been put forward in  explanation, but to date
no general consensus has been reached about a single process
responsible for the spiral structure of galaxies. The most successful
theories employ density waves (Lin \& Shu, 1964) or galaxy interactions
to produce the spiral arms (e.g. Toomre \& Toomre,
1972). Another possibility is stochastic self propagating star
formation considered by Gerola and Seiden (1978). Most probably,
different mechanisms are at work in different galaxies, and these
theories are complementary rather than competing.

On occasion, ejection phenomena have been suggested to be responsible
for spiral arms in galaxies which did not fit into the conventional
schemes. Van der Kruit \etal\ (1972) suggested that the anomalous
radio and emission-line arms in NGC\,4258 originated from an
essentially instantaneous ejection of radio-emitting material from the
centre of the galaxy in opposite directions. In their model, the
observed radio arms are a transient phenomenon and represent the
current position of these plasmons, which were ejected with a wide
range of speeds. The trajectories of the individual parcels is
determined by gravitational forces and ram pressure. Later on, several
other similar models for the jets in NGC\,4258 have been suggested
which are consistent with the observation that the anomalous arms are
bent in the same direction as the trailing spiral arms (e.g. Martin
\etal, 1989, and references therein). 

Wilson \& Ulvestad (1982) described a steady state model in which a
two-sided jet propagates roughly in the rotational plane of a
galaxy. The jet is then bent by the ram pressure exerted transverse
to the jet by the rotating interstellar medium. They applied this
model to the S-shaped radio structures found on a sub-kpc scale in the
Seyfert galaxies NGC\,1068, NGC\,4151, and the radio galaxy
3C\,293. The result of this model are leading spiral structures in the
galaxy. Other mechanisms to obtain S-shaped jet pairs are pressure
gradients in the hot gaseous halo of elliptical galaxies (Smith \&
Norman 1981) and precession of the jets (e.g. Gower \etal\ 1982).

Imaging, both ground based (Hill \etal\ 1988; Steffen \etal\ 1996a,
1996b) and by the Hubble Space Telescope (HST; Capetti \etal\ 1996)
suggests that the spiral structure of the peculiar active galaxy
\iras\ is due to a narrow ridge of emission line filaments. Its
redshift is 0.0462 and was classified as a Seyfert~2 (Beichman \etal,
1985), but the classification has been questioned by Hill \etal\
(1988). It shows an unsually extended and complex radio structure for
this type of galaxies, with a total power of $2.4\cdot10^{23}{\,\rm
W\,Hz^{-1}}$ at 20\,cm. It has extended radio lobes, which are related
to the optical spiral arms. Based on their observations, Hill \etal\
(1988) first suggested that the spiral structure could be due to the
remnant of the interaction of the jets with the ambient medium. In
this scenario the emission line filaments are produced during the
passage of the jet through the interstellar medium. It is then carried
away from its original location by the rotation of the galaxy forming
a spiral structure.

In the present paper we analyze the suggestion by Hill \etal\ (1988)
that the spiral structure in \iras\ is due to the interaction of the
jets with the ambient medium. In Section \ref{hill_mod.sec} we
consider the model proposed by Hill \etal\ (1988). In Section
\ref{bent_jets.sec} we analyze the model proposed by Wilson \&
Ulvestad (1982) for other Seyfert galaxies. Section
\ref{discussion.sec} contains the discussion of our results and our
conclusions are summarized in Section
\ref{conclusions.sec}.

\section{The models}
\label{model.sec}

\subsection{Observational evidence}
\label{evidence}

The main observational reasons for investigating a model 
which invokes the propagation of the jets through the ISM as the
cause for the emission-line spiral structure in \iras\ (Hill \etal\ 1988,
Holloway \etal\ 1996; Steffen \etal\ 1996\,a,b) are, in brief: \\

\noindent
- The spiral structure is emission line dominated. \\
- Radio lobes continue the optical spiral structure. \\
- Radio hotspots and bifurcations of optical spiral arms coincide. \\
- The inner radio double source is aligned (to within $1\,\deg$) 
      with the initial southern hotspot. \\
- The jets themselves are as yet not detected. \\

\noindent
These are very strong arguments for the spiral structure and the radio
ejecta in \iras to be physically associated, rather than related by
coincidence.

\subsection{Straight jets}
\label{hill_mod.sec}

The first three of the crucial observations listed above could
be explained by a bent jet following the spiral structure. However,
the accurate alignment of the inner with the outer radio structure
suggests that the jets are straight.  Therefore, we first 
evaluate the case in which the jets remain unbent during their propagation
through the galaxy.

\begin{figure*}
\centering
 \mbox{\epsfclipon\epsfysize=5in\epsfbox[0 0 356 361]{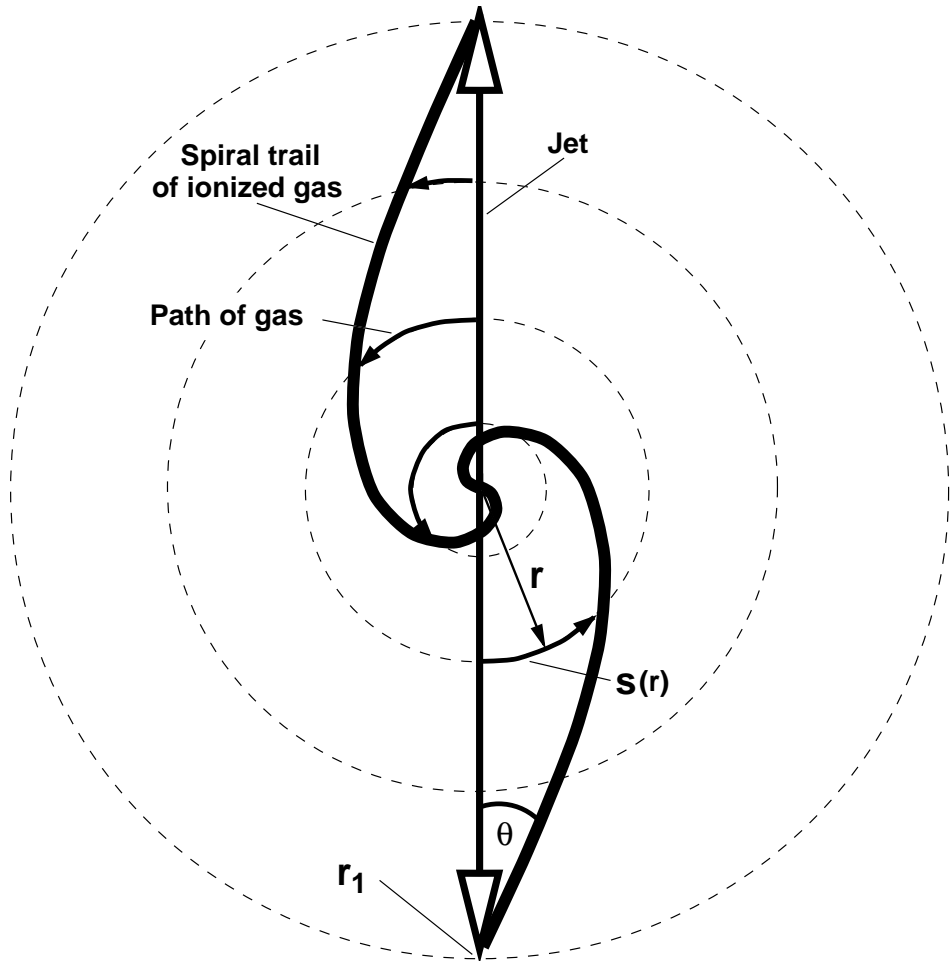}}
\caption{Schematic view of the model with straight jets and
a spiral produced by enhanced density material being dragged away 
from its original position by the rotating interstellar medium.}
\label{spiral_mod.fig}
\end{figure*}

During the passage of a supersonic jet through the interstellar medium
(ISM) the bowshock in front of it sweeps up and compresses the gas in
its path (e.g. Taylor \etal\ 1992). Eventually, after it has moved
away from the bowshock, the gas is entrained by the jet or merges back
into the environment.  What is left in a stationary ambient medium are
filaments of emission-line gas delineating the path of the jet through
its environment. However, in a rotating galaxy the gas can be dragged
away from its original position along an arc of length $s(r)$
following the local motion of the ISM. This situation is illustrated
in Figure~\ref{spiral_mod.fig}, where the head of the jet has reached
a distance $\rc$ since ejection from the centre of the galaxy.

The shape of the spiral arms depends on the rotation curve $v_g(r)$
of the galaxy and on the advance speed $v_{jh}(r)$ of the jet head in
the ambient medium. 

We assume circular orbits of the interstellar medium at a
velocity $v_g(r)$.  The length $s(r)$ of the arc over which the
ionized gas is dragged is then given by

\begin{equation}
s(r) = v_g(r)\int_{r}^{\rc} \frac{dr}{v_{jh}(r)},
\label{arc.eq}
\end{equation}

\noindent
where $r$ is the distance from the centre of the galaxy, and
$\rc$ is the present distance of the jet head, which propagates at
the local speed $v_{jh}(r)$ through the interstellar medium.  
Here and in the following, the subscript `1' indicates the value of a
quantity at the current position of the jet head $r=\rc$. 

An important parameter of the spiral structure related to the advance
speed $v_{jh}$ is the pitch $p$ of the spiral structure close to the
jet head. It can be defined as
\begin{equation}
p_{(\rc)} = \tan\theta = \left.\left(\frac{ds(r)}{dr}\right)^{-1}\right|_{\rc}
  = \frac{v_{jh}(\rc)}{v_g(\rc)},
\label{pitch.eq}
\end{equation}
where $\theta$ is the angle between the jet and the tangent on the spiral 
arm at the position of the jet head (see Figure \ref{spiral_mod.fig}).
If the orientation of the galaxy with respect to the observer is known
or projection effects can be neglected, then this is a measurable
quantity.  As was discussed in Holloway \etal\ (1996), \iras\ is seen
roughly face-on and the jets probably propagate close to the plane of
the sky, we shall therefore ignore projection effects.  The observed
pitch $p$ can therefore be used to estimate the advance speed
$v_{jh}(\rc) \approx p\,v_g(\rc)$ at the position close to the
hotspots (see below).  Other parameters, like the shape of the
rotation curve or a slow change of the jet velocity, have a strong
influence only on the inner spiral structure.  For the simple case of
constant advance speed of the jet head $v_{jh}$, the arc-length $s(r)$
over which the ionized gas at distance $r$ moved around the centre of
the galaxy (while the jet continued to propagate up to a distance
$\rc$) is given by
\begin{equation}
s(r) = \frac{v_g(r)}{v_{jh}} (\rc-r)
\label{arc_lin.eq}
\end{equation}
The advance speed depends mainly on the jet speed $v_j$
and the ratio $\eta$ of the densities of the external medium $\rho_x$
and the jet plasma $\rho_j$ and therefore varies according to the
external conditions. For a highly supersonic non-relativistic jet, as
we assume is the case for \iras, ram pressure arguments lead to a
simple approximate expression for the advance speed $v_{jh}$ of the
jet in an ambient medium (e.g. Leahy 1991)

\begin{equation}
v_{jh} = \frac{v_j}{1+\sqrt{\eta}}
\label{vjh_1.eq}
\end{equation}

To obtain the corresponding relativistic expression, $\eta$ is to be
replaced by $\eta/\gamma^2$, where $\gamma$ is the Lorentz factor
(after Mart\'{\i} \etal\ 1994). Hence, a relativistic jet mimics a heavy
non-relativistic jet and the advance speed approaches the bulk
jet speed $v_j$ as the latter approaches the speed of light. 
Equation \ref{vjh_1.eq} can be combined with Equation \ref{arc.eq}
to obtain the arc-length $s(r)$.

\subsubsection{Evaluation}
\label{evaluation.sec}

We now evaluate the applicability of this model to \iras. The spiral
structure in \iras\ can be roughly characterized by the pitch $p$ of
the spiral at the current position of the jet head. However, the exact
position of the jet head is difficult to determine from the
observations. In our model of a jet crossing a shocked interface
between the interstellar and intergalactic medium, ISM and IGM,
respectively (Holloway \etal, 1996, Steffen \etal, 1996c), it is to be
expected that the expansion speed of the jet material is increased in
the IGM. Therefore, a reasonable choice seems to be an `effective'
position of the jet head shortly beyond the beginning of the hotspot
region. We estimate the pitch of the spiral in \iras\ to be in the
range 1.5\,-\,2.5 and adopt a value of $p=2$. Knowledge of the exact
value is not necessary for our analysis. It is only required that $p$
is of order unity.

Using Equations \ref{arc.eq} \& \ref{vjh_1.eq}, the velocity of the jet
$v_{j1}$ at the position of the jet head can be expressed in terms of
the measurable quantity $p$ and the other parameters as follows

\begin{equation}
v_{j1} = p \cdot v_{g1} (1+\sqrt{\eta_1}). 
\label{v_j_p.eq}
\end{equation}

Extragalactic jets are expected to have very low densities compared to
the ISM, though the range of values is not known.  However,
from the assumption of a straight jet in \iras\ (based on the alignment
of the inner and outer radio structure) and the observed pitch $p$ of
the spiral arms, we can estimate a lower limit for the density ratio
$\eta$.  The radius of curvature $R_j$ of this bending is
approximately given by $R_j/h_j \approx \eta^{-1} v_j^2/v_g^2$, with
$h_j$ representing the jet diameter. The jet speed is unknown. We therefore
use Equation \ref{v_j_p.eq} to eliminate $v_j$ and $v_g$ introducing the
observable pitch $p$ of the spiral arms. 
We find that 
\begin{equation}
\eta = \left(\sqrt{\frac{R_j}{p^2 h_j}} - 1\right)^{-2} 
\end{equation} 
at the position of the jet head.
From the distance $D=4.5$\,arcsec and size $d<0.5$\,arcsec of the
hotspot and a conservative upper limit of the misalignment with the
inner double source ($\alpha < 2\deg$; Steffen \etal\ 1996b), we
estimate that $R_j/h_j \approx D/(2d\sin\alpha) > 130$ and therefore
find the density ratio between the ambient medium and the jet to be
$\eta < 1/22$. From this analysis it is clear that the jet can get
away without being bent by the rotating interstellar medium only if it
is denser than the environment. This conclusion is drawn from the
observations far from the centre of the galaxy. Closer in, at higher
environmental density, the density contrast between the jet and the
ambient medium is required to be even higher. The speed of a heavy jet
is approximately the same as its advance velocity in the ISM. Hence,
in this model, the jet velocity can only be roughly
$v_j \approx p \cdot v_{g1} \approx 400$\kms (using $v_{g1} =
200$\kms).

The high degree of collimation of the jet inferred from the small size
of the hotspots, suggests that the jet is in approximate pressure
equilibrium with the environment.  Then the temperature ratio
between the jet and the environment is roughly the inverse of the
density ratio, and we therefore infer a jet temperature of
\begin{equation}
T_j \approx \eta \cdot T_{ISM} \ll T_{ISM},
\end{equation}
i.e. a very cold jet, possibly with properties similar to galactic jets.

Such a heavy ballistic jet would probably not flare at the interface
between the interstellar and the intergalactic media.  To explain
these observations a collision of the jets with large dense clouds
could be invoked (Higgins \etal\ 1996). However, the symmetry of the
flaring of the two jets on either side, renders this possibility
highly unlikely. From the previous considerations we conclude that the
suggestion by Hill \etal\ (1988) that a pair of straight jets is the
origin of the spiral structure in \iras\ is not a viable model.

\subsection{Bent jets}
\label{bent_jets.sec}

As an alternative, we now analyze a model of jets bent by ram pressure
of the rotating interstellar medium as outlined by Wilson \& Ulvestad
(1982). All assumptions and limitations discussed in their paper 
also applies to \iras. Their main assumptions are: constant jet speed,
adiabatic expansion of the jet, and jet propagation in the plane of
rotation of the galaxy.  As discussed in the previous section, we
shall neglect projection effects. The alignment of the inner double
radio source with the southern hotspot is assumed not to represent the
initial jet direction, as it was the case in the previous section. 

\begin{figure*}
\centering
 \mbox{\epsfclipon\epsfxsize=5in\epsfbox[80 0 611 450]{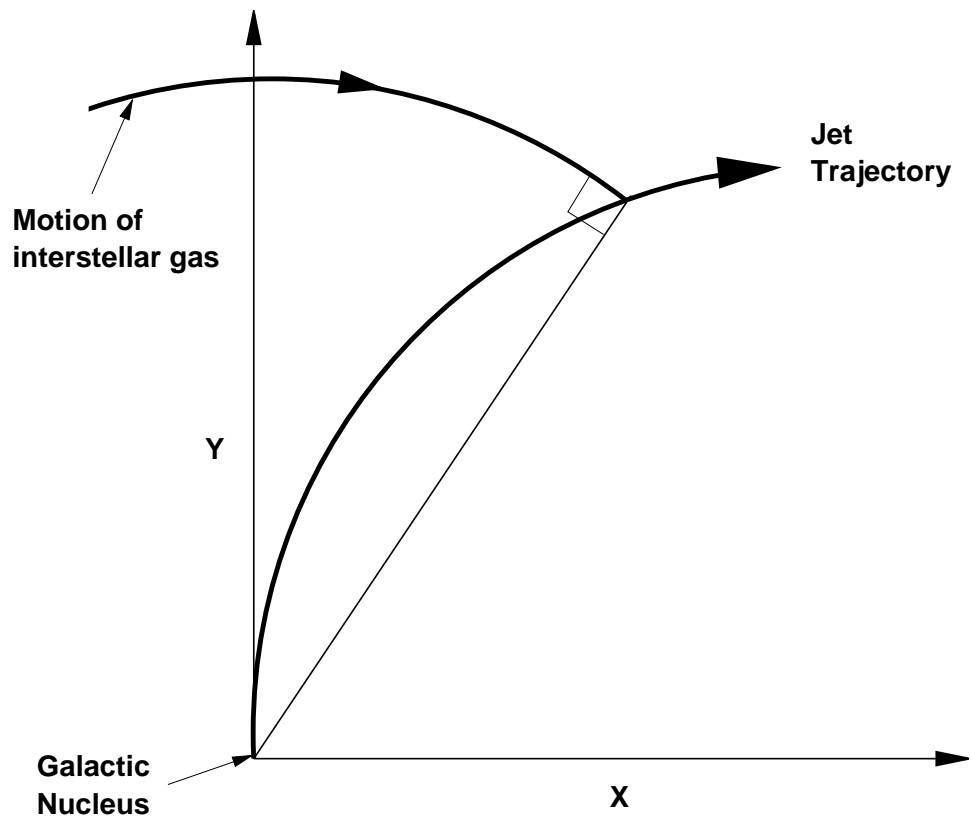}}
\caption{Schematic view of the model with jets bent by the rotating interstellar gas.}
\label{pbend_sc.fig}
\end{figure*}

Figure \ref{pbend_sc.fig} shows a schematic representation of 
the geometry involved in
this model. The steady state path of the jet can be found by numerically 
integrating the equation
\begin{eqnarray}
y''  &=& -\frac{\rho_g(r) \, v^2_g(r)}{\rho_{j0}\, h_0\, v_j^2}
         \left[\frac{\rho_g(r)\, v^2_g(r)}{\rho_g(r)\, v^2_g(r)}
         \right]^{-1/2\Gamma} \nonumber \\
 &&\cdot \left[\frac{(y'y+x)^2}{x^2+y^2}\right]^{(2\Gamma-1)/2\Gamma}
         [1+y'^2]^{(\Gamma+1)/2\Gamma} 
\label{ddx.eq}
\end{eqnarray}

This is Equation (8) in Wilson \& Ulvestad (1982). Here $x$ and $y$ are
Cartesian coordinates, $r$ is the distance
from the galactic nucleus, $\rho$ is the density and $v$ is the
velocity. Subscripts $g$ and $j$ refer to quantities of the galaxy and
the jet, respectively. The effective radius of the jet is $h$, $r_0$
denotes a small starting distance and $\Gamma$ is the adiabatic index.
Initial conditions for the numerical integration are $y=r_0$ and
$y'\gg 1$ at $x=0$.

An important ingredient in Equation \ref{ddx.eq} is the rotation curve
of the galaxy, given by $v_g(r)$.  The rotation curve of \iras\ is
unknown because of the disturbed line profiles and its face-on
orientation with respect to the observer. We considered several
representative rotation curves, which can be described with the
following equation (after Binney \& Tremaine, 1987):
\begin{equation}
v_g(r) =  \frac{r}{a} 
            \frac{v_m}{\sqrt{1+\left( \frac{r}{a}\right)^{\zeta}}}
\label{rot_curve.eq}
\end{equation}
where $a$ is a scale length and $\zeta$ a constant. As representative
examples we choose a rotation curve with a fast transition from solid
body rotation to constant velocity with $a = 0.15$~\kpc\ and a slow
transition with $a=4.5$~\kpc and $\zeta = 2$ in both cases. Choosing
a rotation curve which is slightly rising or falling after the
transition from solid body rotation does not make an appreciable
difference.

\begin{figure*} \centering
  \mbox{\epsfclipon\epsfxsize=2.3in\epsfbox[51 90 519 738]{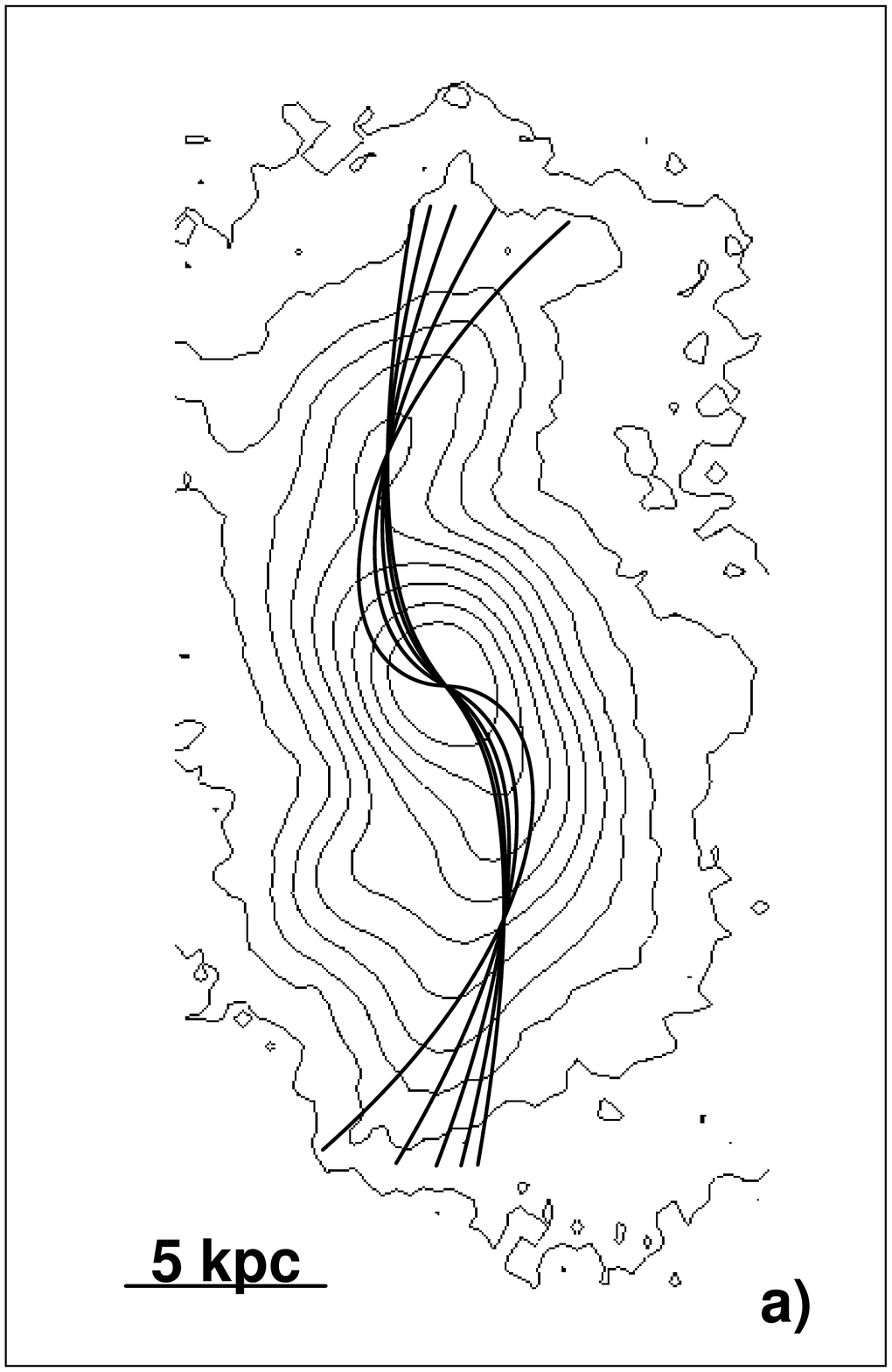}}
  \mbox{\epsfclipon\epsfxsize=2.3in\epsfbox[51 90 519 738]{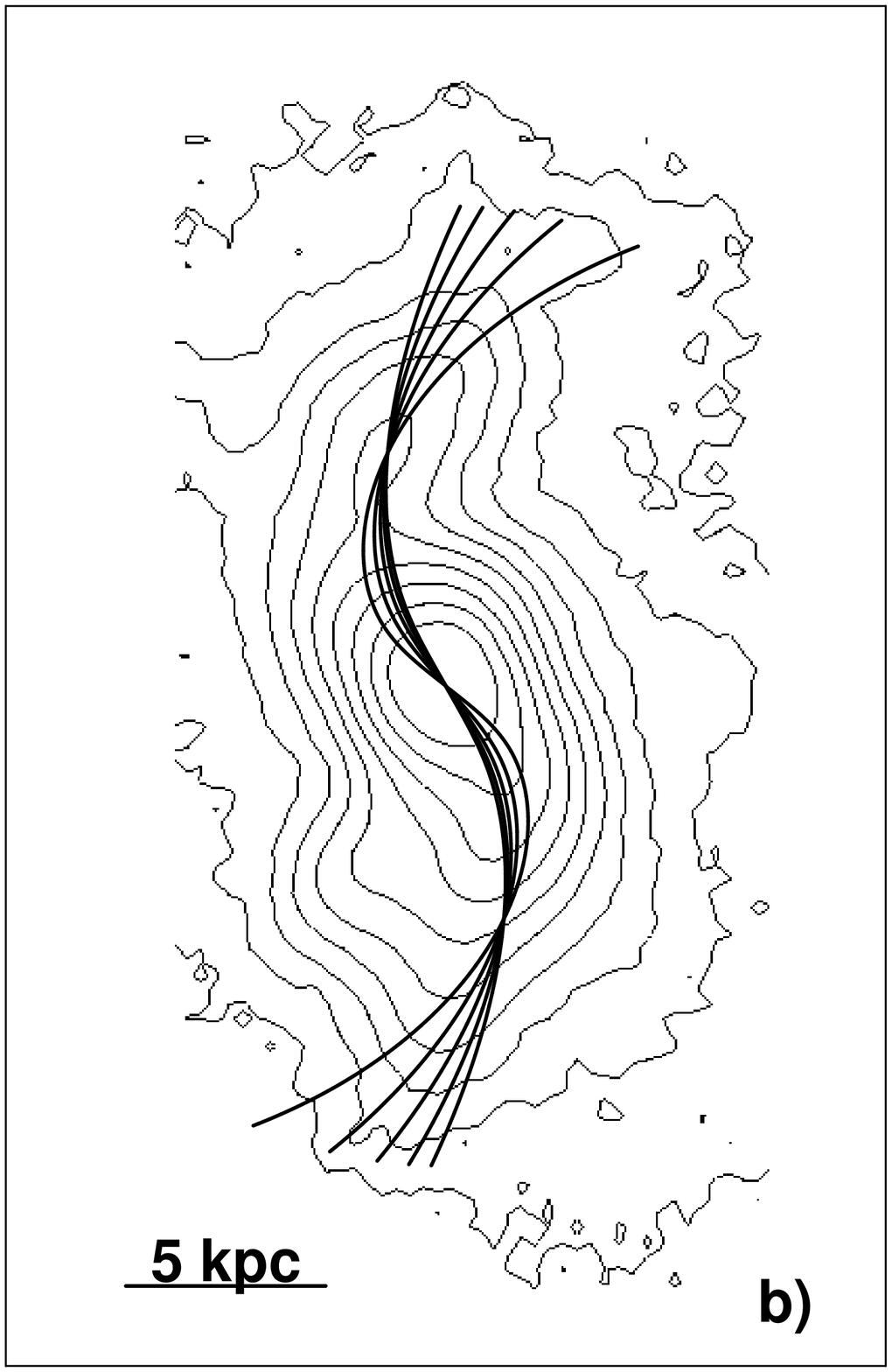}}
  \mbox{\epsfclipon\epsfxsize=2.3in\epsfbox[51 90 519 738]{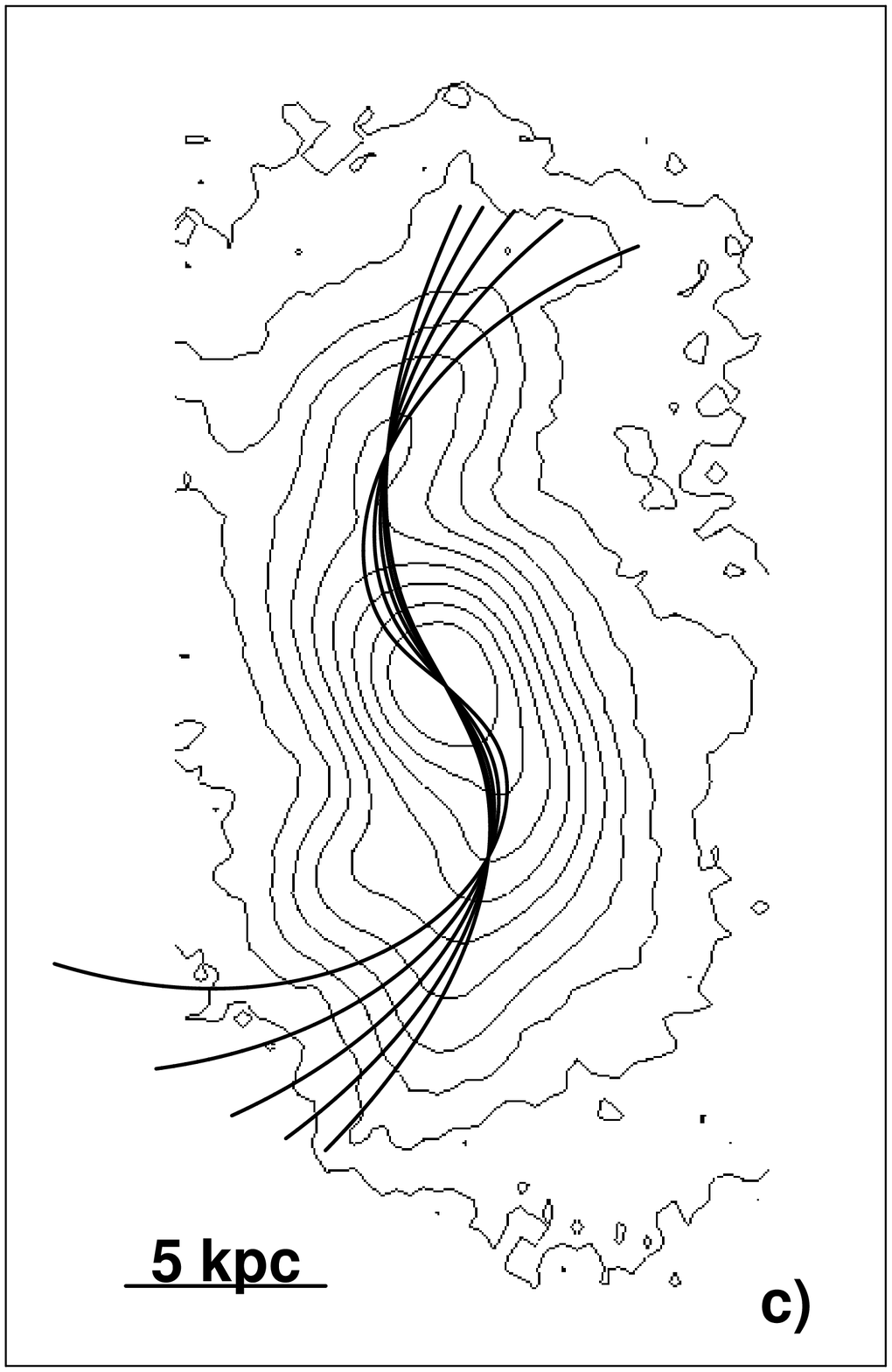}}
 \caption{The observed emission-line spiral structure (Steffen \etal,
 1996c) can be reproduced by a jet bent by the ram-pressure of the
 rotating interstellar medium in the galaxy. The curves in each panel
 show 5 jets with different densities.  Frames `a' and `b' represent
 calculations for different rotation curves, whereas frame `c' is for
 the same rotation curve as `b', but with a higher ambient density in
 the southern spiral arm. The scale of 5~\kpc marked on the frames
 corresponds to 4.5~arcsec (\hubble).}  \label{observe.fig}
 \end{figure*}

Following Wilson \& Ulvestad (1992) we assume that the density of the
galactic ambient medium remains constant within $r < r_d$ and drops
with the square of the distance further out.  Using parameter values
typical for spiral galaxies, in Figure~\ref{observe.fig} we compare
some example trajectories obtained varying the density of the jet.
The values used in the numerical
calculations are given in Table~1. For each of the three examples,
we varied the density of the jet in a range for which the `final'
direction of the jet (arbitrarily chosen at $r\approx 12\kpc$) lays
within the radio lobes.

The individual trajectories have been rotated such that they all pass
through a suitable fitting point on the spiral structure. In the
northern arm we chose the secondary maximum of emission at $r=6$\kpc
(PA$_n=14\deg$). In
Figure~\ref{observe.fig} we compare the results with an \oiii image
which shows the emission line spiral structure of \iras\ (Steffen
\etal\ 1996c). In the frames~\ref{observe.fig}(a) and \ref{observe.fig}(b) the 
calculated spirals are symmetric, using rotation curves `a' and `b',
respectively. They have both been fitted to the northern arm and
clearly expose the asymmetry of the spiral arms in \iras.  In the
southern arm the bending is initially stronger and seems to straighten
out or even `bend back' at $r\approx 5.5\kpc$ (the point of strong
spectral flaring; Holloway \etal\ 1996, Steffen \etal\
1996c). However, the southern arm is brighter and smoother than the
northern equivalent. This could be attributed to a higher galactic
density in this region (rather than lower jet power, because of the
stronger line emission). In Figure~\ref{observe.fig}(c), we therefore
show a specific fit to the southern arm (Case `c') chosing the point
of reference at $r=4.4$\kpc and PA$_s=166\deg$ (the southern tip of
the third contour from the absolute maximum), together with Case `b'
for the northern arm.  Here only difference between the calculations
for the norther and southern arm is an increase of the galactic
density by 30\% in the south, sufficient to obtain the required
bending of the jet.

Note that none of these sets of parameters are unique. Similar results
can be obtained varying the jet velocity or the density of the ambient
medium. Note also that a similar result can be obtained by changing the
density or rotation velocity of the galaxy, as well as the speed of
the jet, since basically only the relative ram-pressure determines the
magnitude of the bending. Shape of the rotation curve
or the density variation as a function of distance do also influence
the jet trajectory to some extent. Although we cannot
determine unique values of the basic parameters involved in the model,
we can at least say that the bent jet model is consistent with the
observations to within reasonable values of the parameters. \\

Within the framework of this model we can now predict where the
underlying jet should point in the inner region of \iras. We suggest
that the northern and the southern radio jets should be oriented at
position angles PA$_n \approx 45\deg$ and PA$_s \approx 135\deg$ with
an estimated error of $10\deg$, as given by the range of initial
position angles listed in Table~1. Within the estimated error, these
values are almost independent of the choice of parameters for the jet
trajectory as long as it satisfactorily fits the spiral structure.

\section{Discussion}
\label{discussion.sec}

We have examined two different candidate models for the formation of
the emission-line spiral structure in \iras\ involving jets ejected
from the galactic nucleus. We exclude the suggestion by Hill
\etal\ (1988) of straight jets which interact with the ISM as a
possible explanation for the observations.  We find that a second
model, considering jets bent by the rotating ambient medium is a
better candidate.

Fitting this model to the optical spiral, we predict that the northern
and the southern radio jets should be oriented along a line inclined
by about $45\,\deg$ to the north-south axis, contrary to the position
angle (6$\,\deg$) of the line joining of the inner radio
components. High sensitivity radio observations will be required to
detect the jets connecting the inner and outer radio structures and to
clarify the nature of the central double source and its alignment with
the southern hotspot.

We found that the initially stronger bending of the southern arm
compared to the northern arm could be attributed to different
densities in the interstellar medium. An asymmetric density
distribution explains not only the initially stronger jet bending in
the southern arm, but also the apparent difference in scale size in
the two arms. In particular, the optical and radio features appear to
be somewhat larger in the northern arm, suggesting a reduced
resistence of the ISM to the action of the jet. The brightness of the
extended northern radio emission is also lower than its southern
counterpart (Steffen \etal, 1996b), another pointer to a reduced
interaction with the environment.  If \iras\ is a disk galaxy, this
effect could be due to a difference in the inclination angle between
the jet and the disk, rather than an intrinsic asymmetry of the
galactic disk (for a discussion of the classification of \iras\ see
Holloway \etal, 1996). However, because of the presence of a close
companion galaxy, even an intrinsic asymmetry due to interaction cannot
be excluded.

The model of a stationary jet bent by ram-pressure provides no direct
explanation for the two strong radio components which are located
within 1~arcsec of the centre of the galaxy. There are broad blue and
red-shifted wings in the spectral lines which have a spatial
separation similar to the radio components, suggesting that they are
related (Holloway \etal, 1996). The longslit spectroscopy and also the
recent HST imaging (Capetti \etal, 1996) shows that the inner
kiloparsec-scale region is very clumpy.  We therefore speculate that
at least the northern one of the central radio components is caused by
the interaction of the jet with a dense cloud, possibly temporarily
cutting off the power supply to the regions further out. This would
explain the reduced brightness of the northern arm (in the optical and
the radio regime) and the northern gap in the emission-line
intensity. Considering the parameters used in our calculations, we
find that this can be achieved by a dense cloud with a diameter of a
few tens parsecs following the galactic rotation at a distance of
1\kpc from the centre (assuming $v_g(1\kpc) \approx 100$\kms).  The
size of the cloud scales with the rotation velocity.

In the present form, the bent jet model assumes that the emission line
gas is located at the position of the jets. However, since it involves
a stationary jet configuration, emission line gas could actually be
dragged away from this path. Therefore, it should show a rather sharp
edge on the upstream and a smoother or filamentary transition on the
side downstream of the environmental gas flow around the jet.

At this time, we cannot exclude jet precession or interaction with the
companion galaxy as origins for the spiral structure in \iras. For
further modeling knowlegde about the rotation of the galaxy is vital,
since the bent jet model predicts that the spiral arms are leading, as
opposed the anomalous arms NGC\,4258, which are trailing, excluding
the stationary ram-pressure bending as a viable model for this galaxy.
Martin \etal\ (1989) suggested that the jet and the tunnel created in
the ISM follows the rotation of the galaxy and is therefore straight
in the corotation zone, while it stays behind further out.  

Detection of collimated jets connecting the inner and outer radio
structure and measurement of the rotation curve of \iras\ would enable
us to distinguish between models with straight jets (e.g. interaction
with its companion or a jet following the corotation) and
non-corotating precession or ram-pressure bending. Since the total
radio power of the galaxy is only $2.4\cdot 10^{23}{\,\rm W\,Hz^{-1}}$
at 20\,cm, very high sensitivity will be required to detect the jets
themselves.  The observation of a displacement between bent radio jets
and the emission-line spiral would indicate that the situation in
reality was some combination between a bent-jet model and the model
discussed in Section \ref{hill_mod.sec}.

To discuss the interaction scenario for \iras\ in any detail, more
spectroscopic information on the companion and a better spectroscopic
coverage of \iras\ itself are needed. Deep imaging of the environment
around \iras\ would be useful to search for traces of the interaction.
Future work should also focus on the classification of the galaxy
(disk or elliptical) and the rotation curve of this unique object.

\section{Conclusions}
\label{conclusions.sec}

We have evaluated two different candidate models for the formation of
the emission-line spiral structure in \iras\ involving jets ejected
from the galactic nucleus. We exclude the suggestion by Hill
\etal\ (1988) of straight jets which interacted with the ISM as a
possible explanation for the observations.  We find that the second
model of two jets bent by the rotating ambient medium is a better
candidate.  Fitting this model to the optical spiral, we predict that
the northern and the southern radio jets should be oriented at
position angles PA$_n \approx 45\deg$ and PA$_s \approx 135\deg$ with
an estimated error of $10\deg$.  Detection of the jets on the
intermediate scale between the inner and outer radio structures will
be necessary to discriminate between the current models. High
sensitivity radio observations will be required to search for the
jets. \\

\section*{Acknowledgements}
AJH and WS acknowledge the receipt of a PPARC studentship and PPARC
research associateship respectively. We thank R.J.R.~Williams and
D.J.~Axon for useful discussions.

\pagebreak

\begin{table*}
\caption{The jet and galaxy parameters for two different rotation 
curves (a,b) are given. For each rotation curve five different jet
densities were considered. For rotation curve `b' two separate sets
(northern and southern arm, respectively) are given, which are
distinguished by the density of the ambient medium. PA is the position
angle of the initial jet direction measured in degrees. Parameters
which are the same for all calculations are: $r_d = 2.2\,\kpc$,
$v_m=200$\,\kms, $r_m$=10\kpc, $r_{j0} = 2$\,kpc, and $h_0 = 3$\,pc.
In the table distances are measured in kpc, velocities in 1000~\kms,
and densities in $10^{-23} {\rm g\,cm^{-3}}$.}

\centering
\begin{tabular}{lllll}
\\
\hline
case & $a$ & $\rho_{g0}$ & $ \rho_{j0}$ & PA 
\vspace*{1mm} \\
a  & 0.15 & 30. & 1.0 & 92 \\
   & 0.15 & 30. & 1.5 & 68 \\
   & 0.15 & 30. & 2.0 & 55 \\
   & 0.15 & 30. & 2.5 & 48 \\
   & 0.15 & 30. & 3.0 & 43 
\vspace*{1mm}  \\
     b & 4.5 & 30. & 0.06 & 58 \\
north  & 4.5 & 30. & 0.08 & 48 \\
       & 4.5 & 30. & 0.12 & 42 \\
       & 4.5 & 30. & 0.12 & 38 \\
       & 4.5 & 30. & 0.14 & 35
\vspace*{1mm}  \\
     b & 4.5 & 40. & 0.06 & 127 \\
south  & 4.5 & 40. & 0.08 & 135 \\
       & 4.5 & 40. & 0.10 & 141 \\
       & 4.5 & 40. & 0.12 & 144 \\
       & 4.5 & 40. & 0.14 & 147 \\
\hline
\end{tabular}
\end{table*}

 \begin{figure*}
 \centering
 \mbox{\epsfclipon\epsfxsize=5in\epsfbox[72 230 540 570]{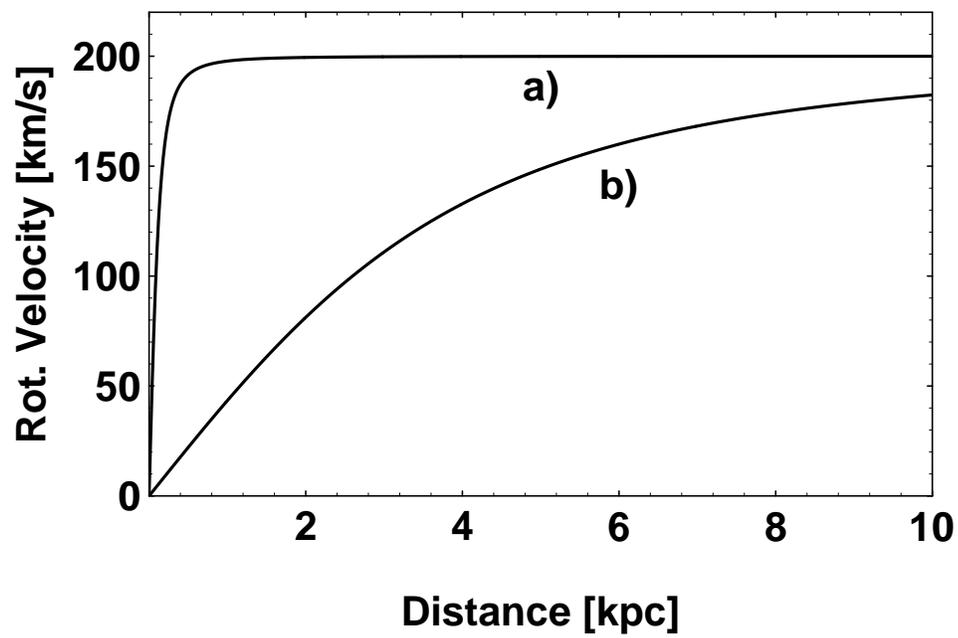}}
 \caption{Rotation curves with a sharp and a slow rise have been considered.}
 \label{rotcvs.fig}
 \end{figure*}

\end{document}